\documentclass[a4paper]{article}
\usepackage[utf8]{inputenc}
\usepackage{hyperref}
\usepackage{csquotes}
\usepackage{indentfirst}
\MakeOuterQuote{"}

\title{Democratizing online controlled experiments at Booking.com}
\author{Raphael Lopez Kaufman \\ raphael.lopez@booking.com
	\and Jegar Pitchforth \\ jegar.pitchforth@booking.com
	\and Lukas Vermeer \\ lukas.vermeer@booking.com}
\date{}

\begin{document}

\maketitle

\begin{abstract}

There is an extensive literature about online controlled experiments, both on the statistical methods available to analyze experiment results \cite{Rubin1974,Box2005,Xie2015} as well as on the infrastructure built by several large scale Internet companies \cite{Tang10,Kohavi2013, Xu, Bakshy14} but also on the organizational challenges of embracing online experiments to inform product development \cite{Xu, Fabijan2017}. At Booking.com we have been conducting evidenced based product development using online experiments for more than ten years. Our methods and infrastructure were designed from their inception to reflect Booking.com culture, that is, with democratization and decentralization of experimentation and decision making in mind.

In this paper we explain how building a central repository of successes and failures to allow for knowledge sharing, having a generic and extensible code library which enforces a loose coupling between experimentation and business logic, monitoring closely and transparently the quality and the reliability of the data gathering pipelines to build trust in the experimentation infrastructure, and putting in place safeguards to enable anyone to have end to end ownership of their experiments have allowed such a large organization as Booking.com to truly and successfully democratize experimentation.

\end{abstract}

\section*{Introduction}

At Booking.com we have been using online controlled experiments for more than ten years to conduct evidence based product development. Overall, on a daily basis, all members of our departments run and analyse more than a thousand concurrent experiments to quickly validate new ideas. These experiments run across all our products, from mobile apps and tools used by hoteliers to customer service phone lines and internal systems.

Experimentation has become so ingrained in Booking.com culture that every change, from entire redesigns and infrastructure changes to bug fixes, is wrapped in an experiment. Moreover, experiments are used for asynchronous feature release and as a safety net, increasing the overall velocity of the product organization. Finally, they are also a way of gathering learnings on customer behaviour, for example by revisiting previous successes and failures.

It is with the aim of allowing such ubiquitous use cases, and of truly democratizing experimentation, that we built an in-house experiment infrastructure trustworthy and flexible enough to lead all the products in the right direction.  Indeed, not only does each department have its own dedicated team to provide support and to work on improving the experiment platform, but all the steps required to validate new product ideas are fully decentralized: gathering learnings from previous experiments to inform new hypotheses, implementing and running new tests, creating metrics to support new areas of the business and most importantly analysing the results and making decisions. Such democratization is only possible if running experiments is cheap, safe and easy enough that anyone can go ahead with testing new ideas, which in turn means that the experiment infrastructure must be generic, flexible and extensible enough to support all current and future use cases.

In many respects the components of Booking.com experiment infrastructure are similar to what Microsoft, Linkedin or Facebook have described in previous contributions \cite{Kohavi2013, Xu, Bakshy14}. However, in this paper, we share some key features of our infrastructure which have enabled us to truly democratize and decentralize experimentation across all the departments and products at Booking.com. 

\section*{Central repository of successes and failures}

Enabling everyone to come up with new hypotheses is key to democratizing experimentation and move away from a product organization where only product managers decide what features to test next. Therefore, Booking.com experiment platform acts as a searchable repository of all the previous successes and failures, dating back to the very first experiment, which everyone can consult and audit. Experiments can be grouped by teams, areas of the Booking.com products they ran on, segments of visitors they targeted and much more. The data of these previous experiments is shown in the exact same state as it was to their owners, along with the hypotheses that were put to the test. Moreover, descriptions of all the iterations and of the final decision, whether the experiment was a success or not, are available.

Having such a consistent history implies the extra work of making backwards compatible changes to the reporting platform and of keeping all the experiment data, but it has proven immensely worthwhile as cross pollination between teams and products, by iterating on or revisiting past failures and by disseminating successes led to many improvements for our customers. However, one of the challenges we are still facing is being able to answer questions such as "what were the findings related to improving the clarity of the cancellation policies in the past year" in the central repository, to go beyond short term cross team and cross departments learnings, without resorting to using a set of tags to label experiments. Indeed, this approach is proving not sustainable both because it involves making sure the tags are normalized and kept up to date and because it entails poor discoverability and usability of the central repository. 

\section*{Genericity and extensibility}

To support experimentation across several departments and several products, the platform needs to be generic enough to allow the setup, implementation and analysis of an experiment with minimal ad hoc work, apart from the implementation of new business logic. That means recruitment, randomization and recording visitors' behaviour is abstracted away behind a set of APIs made available for all products. Reporting is also automatically handled by the infrastructure and is department and product agnostic. However, both the API and the reporting need to be easily extensible to support as many use cases as possible.

Firstly, we provide an extensible metric framework. Anyone can create new metrics, may they be aggregated in real-time or using batch updates, depending on what is more suitable. One can choose to have these metrics automatically aggregated for every experiment because of their general business scope or to make them available on demand. Finally, the framework accommodates for both binary (e.g. is this visitor a booker) and real valued metrics (e.g. what is the euro commission of a booking).

Secondly, to support cross products experimentation and experiments on new products we also provide an extensible framework to identify visitors, called tracking methods. Example of such tracking methods are user account and email address based tracking. Indeed, oftentimes the effects of an experiment are not limited to one product. Changing the way customers interact with our mobile apps is likely to also change the way they use our main website.

As an example, consider even the simplest experiment which aims at increasing bookings by changing the way hotel pictures are displayed in the Android app. It is likely that many people will consult the Booking.com website on their phone during their commute to decide on a hotel but will eventually book the trip on their desktop at home. Therefore, we cannot only identify visitors using their http cookie, which is tied to a specific browser on a single device, if we are to truly assess the impact on our customers of this new idea, but we must find a way to identify visitors which is consistent across their journey on our different products. Adding a new tracking method can be done by any developer and we now have more than a dozen ways to identify visitors across all our products. 

\section*{Data which can be trusted}

New hires coming from more traditional product organizations often find themselves humbled and frustrated when their ideas are invalidated by experiments. As such, building trust in our infrastructure is key to democratizing experimentation. It means that experiment results must draw as accurate a picture as possible of the true effect of a new feature on our customers. It is also a necessary condition to ensure that decentralized decision making leads to customer experience improvements and, eventually, business growth.

To address this, Booking.com experiment infrastructure teams have made data quality a priority. We monitor the validity of the data used for decision making by computing a set of common metrics in two entirely separate data pipelines maintained by different engineers who do not review each other's code, one doing near real-time aggregations (less than a five minutes' delay) and one doing daily batch updates. This allows us to quickly detect bugs in the aggregation code when we ship new features to these pipelines, and to strengthen our trust in the quality of the data they generate by alerting on the discrepancies we found for these metrics on real experiments \cite{Silberzahn2017}.

We rely on monitoring real experiment data rather than on a simple test suite because discrepancies can arise not only because of code changes in the aggregation pipelines but because of code introduced by other teams. As an example consider how cancellations are recorded. Not only do we need to keep track of cancellations done by customer themselves but also by customer support agents and even properties owners and hotel managers. The two pipelines use different data sources for cancellations corresponding to the different needs of real-time and batch processing. Any change (e.g. adding a new endpoint for hoteliers to do bulk cancellations) which would lead to cancellations events being fed to one of the sources but not to the other will result in a discrepancy in the cancellation count between the two pipelines.

Moreover, to make sure that no piece of information is lost between when it is first generated on our products to when it is received by our aggregation pipeline, we maintain a set of experiments for which we control the input (e.g. number of visits). Finally, we also monitor the statistical framework used for decision making by maintaining a pool of AAs (experiments whose treatment does not introduce any sort of change) which allow to validate its theoretical properties (e.g. false positive rates or the statistical distribution of metrics).

This complex monitoring infrastructure is also maintained to foster a critical thinking attitude towards data among experiment owners. Indeed, all too often, data is taken as the absolute source of truth. Therefore, in the reporting itself it is possible to see by how much the two aggregation pipelines diverge, because of inevitable data loss for example. Similarly, the pool of AAs exemplifies the concept of a false positive, which is at first hard to grasp.

\section*{Loose coupling between the experiment infrastructure and the business logic}

Making sure that the pipelines consistently aggregate data is, however, not sufficient. Indeed, for experimentation to allow successful evidence based product development we must ensure that whatever results are reported to experiment owners are an accurate measurement of the true impact of their new features on the customers who would be exposed to the changes were they deemed successful enough to be permanently shipped. Therefore, it was decided very early to keep the experiment infrastructure as loosely coupled to the business logic as possible.

In our infrastructure, the target of an experiment (e.g. logged in, English speaking and looking for a family trip customers) and where the experiment runs (e.g. in the hotel descriptions in the search result page) is implemented in the code by experiment owners and is not addressed by the experiment platform. It may seem like a decrease in velocity since new code needs to be written, tested and rolled out. However, this decrease is offset by the fact that exposing to the new feature all the types of customers that were targeted during the experiment runtime, and only those, is just matter of removing the API call that was used to decide treatment assignment. Indeed, let us imagine that the infrastructure responsible for recruiting visitors in experiments would have to be aware of segments (e.g. leisure or business). Once an experiment is successful, that would mean that either the line of code which does visitor recruitment needs to stay forever, adding to the complexity of the codebase (indeed, static analysers doing cleanups would also have to be aware of experiment statuses), or that cleaning the experiment logic involves adding a line of code which was never tested and is not guaranteed to use the same definition of segments.

Moreover, experiment state data is not exposed via an API, purposefully. For example, it is not possible to know, in the business code, whether a given experiment is running or in which experiments a visitor was recruited. Indeed, this would lead experiment results to be context dependent (e.g. on the set of experiments which are running concurrently) and therefore be poor predictor of the future business impact of new features.

\section*{Building safeguards}

When anyone is empowered to run new experiments, it is crucial to have a very tight integration between the experiment and the monitoring platforms. Sometimes experiments introduce such severe bugs, removing for example the ability for certain customers to book, that the overall business is immediately impacted. Therefore, it must be possible to attribute, in real time (in our current system that means less than a minute), the impact of every experiment on both the overall business and the health of the systems.

However, having experiment owners closely monitor their experiments is not enough. Everyone in the company must be able and feel empowered to stop harmful experiments of their own accord. It is therefore as much a technical as a cultural issue. It is worth noting that given our infrastructure, we could easily automate such stops. However, we feel this goes contrary to democratizing experimentation which fosters end to end ownership. Moreover, deciding whether an experiment is more harmful than beneficial, given that some experiments are aiming at gaining a better understanding of visitors' behaviour rather than bringing a "win", is a judgement call which is best made by those who own the change.

Safeguards must also be established when it comes to decision making. Indeed, experiment results must accurately predict the future impact on customers of new features. Given the frequentist statistical framework we use at Booking.com, that means we need to enforce good practices as much as possible to make sure successes are not statistical ghosts.

One such important practice is pre-registration. Experiment owners need to specify up front which customer behavior they want to impact and how, the set of metrics which is going to support their hypothesis, and how these metrics are going to change. Coupled with a culture which encourages peer review of successes this enforcement considerably reduces p-hacking \cite{Munafo2017} and dubious decisions.

We have also implemented safeguards regarding missing data. For experiments relying on asynchronous or lossy channels for tracking, such as events generated from Javascript code, statistical checks are in place to detect selective attrition (a phenomenon which renders all experiment results void \cite{Zhou2016}). In that case, not only a warning is visible on the experiment report but comparative statistics used for decision making are hidden, making sure no oblivious experiment owner can disregard the issue.

As an example, consider an experiment which loads synchronously additional images. Visitors with slow internet connection will drop off at a higher rate in treatment than in control before any data can be sent back to our servers, leading to selective attrition. Similarly, whenever the experiment API is called on a visitor whose identifier is unknown (e.g. when an experiment is targeting logged in users and tries to recruit a non logged in visitor), the visitor is shown control and is not recruited in the experiment, and their behaviour is not recorded. In such cases a message is displayed on the reporting itself to warn experiment owners that the data they are looking at is incomplete. Indeed, the data shown in the reporting might not be an accurate representation of the true business impact of the change anymore, as the visitor whom we could not identify would be exposed to the change were it to be permanently shipped. However, in this case, contrary to selective attrition (missing data not at random), full ownership of the decision is left to experiment owners rather than relying on automation. Indeed, many factors are to be taken into account when assessing the impact of missing data at random on experiment result validity. For example, the amount of missing identifiers can be insignificant compared to the magnitude of the improvement brought by the treatment, or the missing identifiers may be caused by malicious traffic to our products with, therefore, no impact on the accuracy of the data reported on the business impact of the experiment.

\section*{Conclusion}

Enabling anyone to form new hypotheses by maintaining a repository of past failures and successes, building trust in experimentation by accurately measuring customer behaviour, making experimentation accessible and safe by designing extensible frameworks with built-in safeguards, these are the keys features of our experiment infrastructure that we hope will help other companies truly democratize experimentation in their product organization. Moreover, we showed how much emphasis was put on giving as much ownership as possible to experiment owners. This entails that almost everyone at Booking.com needs to be aware of many of the inner workings of experimentation, regarding, for example, hypothesis testing, data collection and metrics implementation, in order to be able to make the best informed judgement calls on their experiments. Therefore, dedicating considerable time to trainings, both online and in classrooms, and to in person support is necessary for decentralizing experimentation at scale. 

However, we are still facing many unanswered challenges to make sure anyone at Booking.com can leverage online controlled experiments to improve customer experience: designing cross-department metrics, exploring the tradeoffs between velocity and accurate business impact measurement, meaningful assessment of server-side performance related experiments.

\end{document}